# Development of Large Area GEM Chambers

J. Yu[1], E. Baldelomar[1], K.Park[2], S. Park[1], M. Sosebee[1], N. Tran[1], A.P.White[1]

[1]*Dept. of Physics, University of Texas at Arlington, Arlington, Texas, USA*
[2]*Korea Atomic Energy Research Institute, Daejeon, Korea*

**Abstract**

The High Energy Physics group of the University of Texas at Arlington Physics Department has been developing Gas Electron Multiplier (GEM) detectors for use as the sensitive gap detector in digital hadron calorimeters (DHCAL) for the future International Linear Collider. In this study, two kinds of prototype GEM detectors have been tested. One has 30x30 $cm^2$ active area double GEM structure with a 3 mm drift gap, a 1 mm transfer gap and a 1 mm induction gap. The other one has two 2x2 $cm^2$ GEM foils in the amplifier stage with a 5 mm drift gap, a 2 mm transfer gap and a 1 mm induction gap. We present characteristics of these detectors obtained using high-energy charged particles, cosmic ray muons and $^{106}$Ru and $^{55}$Fe radioactive sources. From the $^{55}$Fe tests, we observed two well-separated X-ray emission peaks and measured the chamber gain to be over 6500 with a high voltage of 395 V across each GEM electrode. Both the spectra from cosmic rays and the $^{106}$Ru fit well to Landau distributions as expected from minimum ionizing particles. We also present the chamber performance after high dosage exposure to radiation as well as the pressure dependence of the gain and correction factors. Finally, we discuss the quality test results of the first set of large scale GEM foils and discuss progress and future plans for constructing large scale (100cmx100cm) GEM detectors.

## 1. INTRODUCTION

Gas Electron Multiplier (GEM)[1, 2] technology was invented at CERN late 1990s for high precision fast tracking use. Given its intrinsic position resolution, it is a detector technology that can be used in a high granularity calorimeter, as long as the Landau tail of the responses can be controlled. In addition, a large scale (over 1mx1m) GEM chambers need to be constructed to cover the large area needed for calorimeters. For use in a calorimeter which covers a much larger area to fill the entire detector volume, it is of critical importance to be able to construct large area GEM detectors with minimal dead area. In addition, the large area GEM detectors are useful for much broader areas beyond high energy physics detectors, such as a whole body medical imaging device, homeland security devices or space based astrophysics experiment telescopes.

Over the past several years the University of Texas at Arlington (UTA) team and collaborators have been developing Digital Hadron Calorimetry (DHCAL) [3 – 7] using GEM as the sensitive gap detector technology. DHCAL is a solution for allowing a particle flow algorithm (PFA) [8] to be used in precision jet energy measurement since GEM can have flexible configuration which allows small anode pads for high granularity. It is robust and fast with only a few nano-second rise-time, and has a short recovery time, which allows higher rate capability than other detectors, such as a resistive plate chamber (RPC). It operates at a relatively low voltage across the amplification layer, can provide high gain using a simple gas ($ArCO_2$) which protects the detector from long term issues, and is stable.

UTA's GEM prototype chamber ionization signal from charged tracks passing through the drift section of the active layer is amplified using a double GEM layer structure. The amplified charge is collected at the anode layer with $1\,\text{cm}\times 1\,\text{cm}$ pads at zero volts. The potential differences to guide the ionization electrons are produced by a resistor network, with successive connections to the cathode, both sides of each GEM foil, and the anode layer. The pad signal is amplified, discriminated, and a digital output produced. GEM design allows a high degree of flexibility with, for instance, possibilities for micro-strips for precision tracking layer(s), variable pad sizes, and optional initial ganging of pads for eventual finer granularity future readout if required and allowed by cost considerations.

## 2. The Initial Study with $10\,\mathrm{cm}\times 10\,\mathrm{cm}$ and $30\,\mathrm{cm}\times 30\,\mathrm{cm}$ Prototype Chambers

The initial studies were conducted on signal characteristics and gain from a prototype GEM detector using $10\,\mathrm{cm}\times 10\,\mathrm{cm}$ GEM foils. The signal from the chamber was read out using the QPA02 chip developed by Fermilab for Silicon Strip Detectors. The gain of the chamber was determined to be of the order 3,500, consistent with measurements done by the CERN GDD group. The MIP efficiency was measured to be 94.6% for a 40 mV threshold, which agrees with a simulation of chamber performance. The corresponding hit multiplicity for the same threshold was measured to be 1.27, which will be beneficial for track following and cluster definition in a final calorimeter system. A gas mixture of 80% Ar/20% $CO_2$ has been shown to work well and give an increase in gain of a factor of 3 over the 70% Ar/30% $CO_2$ mixture. A minimum MIP signal size of 10 fC and an average size of 50 fC were observed from the use of this new mixture. The prototype system has proved very stable in operation over many months, even after deliberate disassembly and rebuilding, returning always to the same measured characteristics. We investigated cross talk properties using nine $1\,\mathrm{cm}\times 1\,\mathrm{cm}$ cell anode pad layout. We also used collimated gamma rays from a $^{137}$Cs source to study signal sharing between adjacent pads. These chambers were read out using the 32-channel QPA02 chip based Fermilab preamp cards. We conducted three beam tests to measure the rate capability of the chamber, its MIP characteristics, cross talk between the channels and occupancy. The output signals from the amplifier cards were sent to discriminator boards which contain discriminator chips, multiplexer stages, and data output interface. The output from the discriminator boards were read out by a PCI based ADLink ADC controlled by LabView software.

The first beam exposure of our $30\,\mathrm{cm}\times 30\,\mathrm{cm}$ prototype chamber took place in May 2006 at the high flux beam which consists of 30 ps pulses of $10^{10}$ electrons every 43 μs in 5 cm radius; the detector and the electronics measured responses to $10^9$ electrons per pad. While the electronics was saturated, the chamber was able to see the beam clearly and provided a good measure of the time structure of the beam. As a test, we directly exposed a broken GEM foil to the beam. In both the chamber and the broken GEM foil, we did not see any physical damage. In addition, while the signal shapes were distorted by the hit from $10^9$ electrons per pad, the chamber responded well to such a large signal, giving us confidence that the chamber will function in the ILC environment without damage.

**Figure 1. (a) GEM DHCAL concept diagram with a double GEM layers as shown in the inset to keep the sensitive gap size as small as possible (b) UTA GEM prototype chamber constructed with $30\mathrm{cm}\times 30\mathrm{cm}$ CERN GDD GEM with 64 channel KPiX readout board**

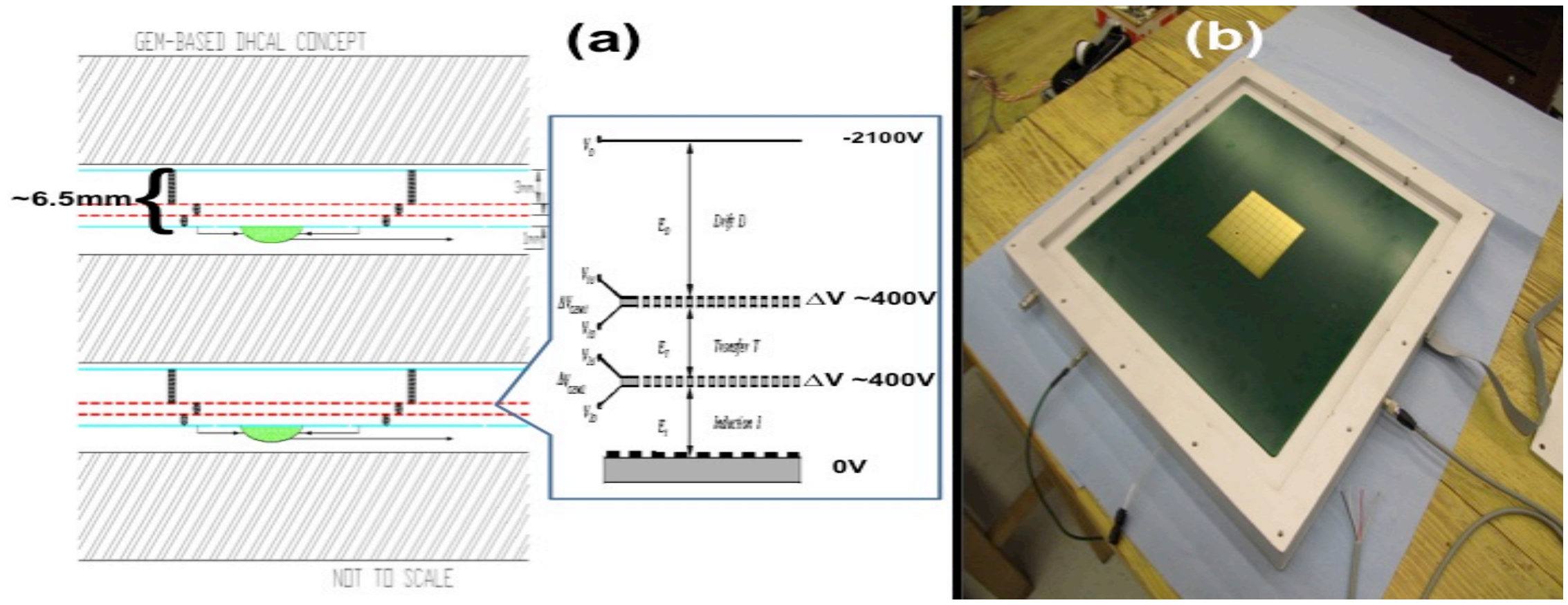

Additional beam tests were conducted at Fermilab's Meson Test Beam Facility (MTBF) [9] in April 2007. Most the useful data taking was done using 120 GeV proton beams from the Main Injector. Labview based online analysis software complimented the DAQ software and allowed us to monitor the data as they were accumulated.

## 3. Development and Characterization of $30\,\mathrm{cm}\times 30\,\mathrm{cm}$ Prototype Chambers using 13-bit KPiX Chip

As the continuation of the chamber characterization process, and as the first step for the full-size ( $1\,\mathrm{m}\times 1\,\mathrm{m}$ ) test beam chambers, we have been developing a $30\,\mathrm{cm}\times 30\,\mathrm{cm}$ GEM prototype chambers read out with the 64 channel 13-bit KPiX[10] chip originally developed for the SiD silicon-tungsten (Si/W) electromagnetic calorimeter (ECAL) [11], [12]. The chip has been modified to include a switchable gain to accommodate small signals from GEM chamber. Fig.1.(b) shows a photo of the readout anode board integrated into a prototype chamber.

For mechanical assembly, we have developed G10 based grid spacers with the grid size $8\,\mathrm{cm}\times 8\,\mathrm{cm}$ sufficient for keeping the foils flat and maintain efficient uniform gas flow throughout the entire detector gas volume. The detector enclosure was constructed with a one-piece aluminium block for noise shielding and gas tightness. We constructed several $30\,\mathrm{cm}\times 30\,\mathrm{cm}$ prototype chambers using the foils manufactured by the CERN GDD group despite the initial development of the foils with 3M Inc. due to the fact that the company sold their flex circuit division.

We helped in characterizing a couple of development versions of KPiX chips with GEM chamber to verify the functionality of the chip and the integrated detector system. For this study, we have regularly taken calibration data hourly to see if there are day – night and weekday – weekend effects. This is to fully understand whether there are environmental effects that would impact our measurements with KPiX in our labs due to varying noise characteristics. We observed that the mean values of any given KPiX channel do not show any systematic day – night dependence or weekday – weekend dependence. The fluctuation in pedestal mean value for each channel is within 3 – 5%. We, however, observe the mean value of the pedestal varies between 20 and 130 ADC counts channel to channel. We also observed that the channel-to-channel variation of the gains vary 5 – 20 ADC counts/fC. These observations have been dramatically improved in newer version of the chip to work with our prototype GEM detectors.

**Figure 2. (a) Cosmic ray test stand with KPiX DAQ electronics. Inset shows three $30\,\mathrm{cm}\times 30\,\mathrm{cm}$ prototype chambers of which one is read out with the 13-bit KPiX chip while the other two with 1-bit DCAL chip. (b) A schematic diagram of one of the 64 channels of the 13-bit KPiX chip.**

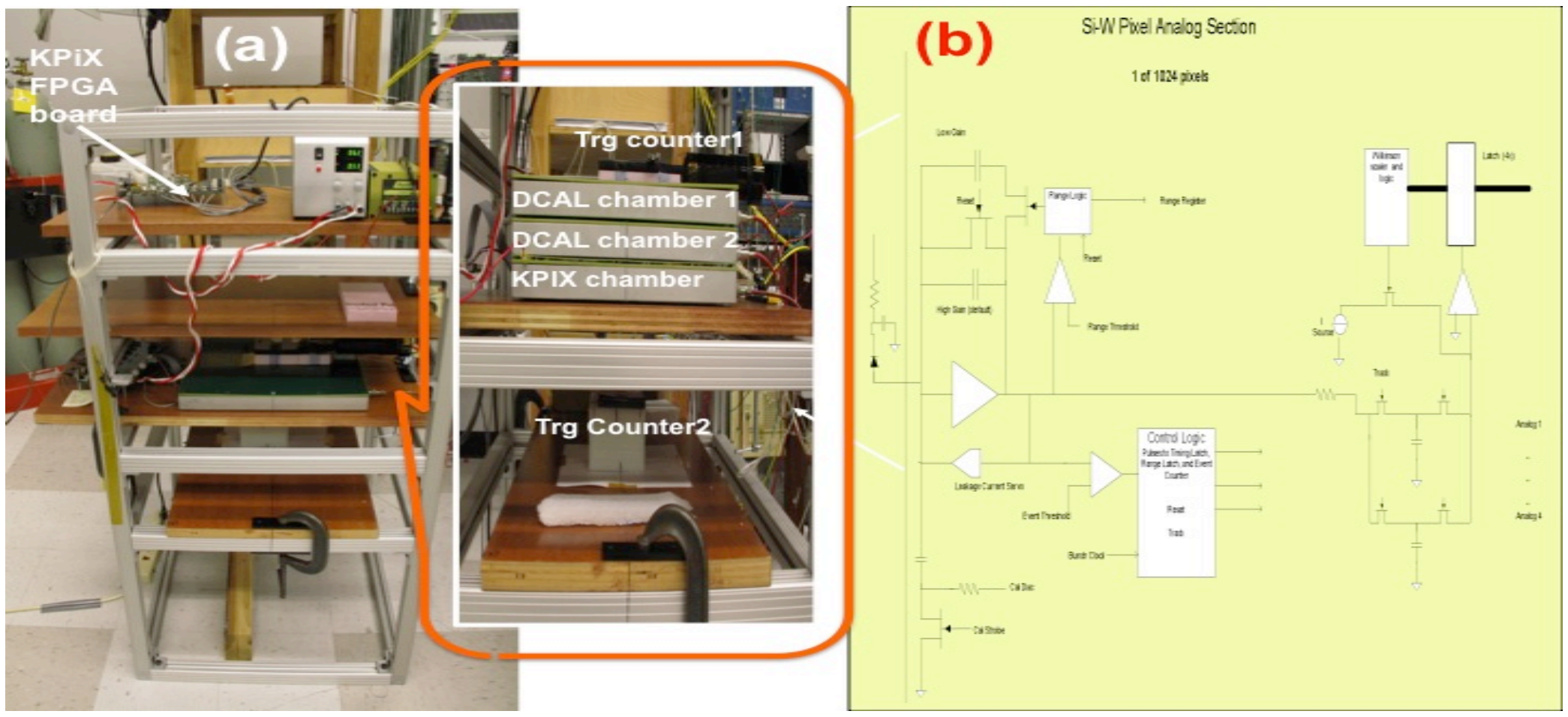

The next generation prototype 30cmx30cm GEM detector prototype was integrated with the KPiX V7 chip. We characterized the prototype detector on the bench using various radioactive sources, including $^{55}$Fe X-ray source, $^{106}$Ru, Sr$^{90}$ and cosmic rays. Our prototype chambers use $ArCO_2$ mixture in 80/20 proportion. Figure 2.(a) shows our cosmic ray test stand with KPiX, FPGA and interface board. The inset shows the chamber integrated with the KPiX chip while Fig. 2.(b) shows a schematic circuit diagram of the KPiX chip which clearly shows the switchable gain and integrated trigger logic. This new version of the KPiX chip was modified to accept random triggers (at the cost of duty factors) to compensate the originally designed accelerator based clock from the International Linear Collider (ILC), allowing the triggers for cosmic ray and beam tests.

As in the case for a gas detector system that releases the output gas to the atmosphere, UTA GEM prototype detectors' gains depend on the atmospheric pressure variations. To correct for the pressure variations, we obtain the correction factor as a function of atmospheric pressure and correct all gains to that of 1atm. Figure 3.(a) shows the correction factor which shows a linear dependence of the gains as a function of the atmospheric pressure. Figure 3.(b) shows the pressure corrected charge distributions of β-rays from $^{106}$Ru source while Fig.3(c) shows the characteristic signal distribution of $^{55}$Fe source.

Finally, in the process of constructing prototype chambers using the 30cmx30cm foils, we have encountered gains smaller than anticipated. We have learned that the standard foil production technique at present focuses more on the stability of the detector than the gains. Since for DHCAL, it is necessary to maintain the detector width as small as possible, we use double-GEM layer configuration instead of triple-GEMs. This requires the gains from each of the GEM layer to be as high as possible without damaging the foil surfaces from frequent discharges. To remedy this issue, we requested CERN to produce GEM foils with higher gain technique and measured the gains from detectors with these two different sets of GEM foils. Figure 3.(d) shows the gain as a function of the high voltage across the entire detector (the voltage across each GEM foils is 1/5 of this voltage). It is evident that the two sets of GEM foils show dramatic differences in gains. Based on this measurement, we have decided to construct 30cmx30cm prototype detectors for beam test using the higher gain GEM foils.

**Figure 3. (a) Pressure correction for GEM gains (b) MIP distributions of β-rays from $^{106}$Ru (c) Characteristic charge distribution from $^{55}$Fe X-rays (d) gains as a function of HV across the entire chamber (each GEM foils has 1/5 of this voltage across it) for hole shape dependence. The foil with smaller gap size has higher electric field line density, causing higher gains.**

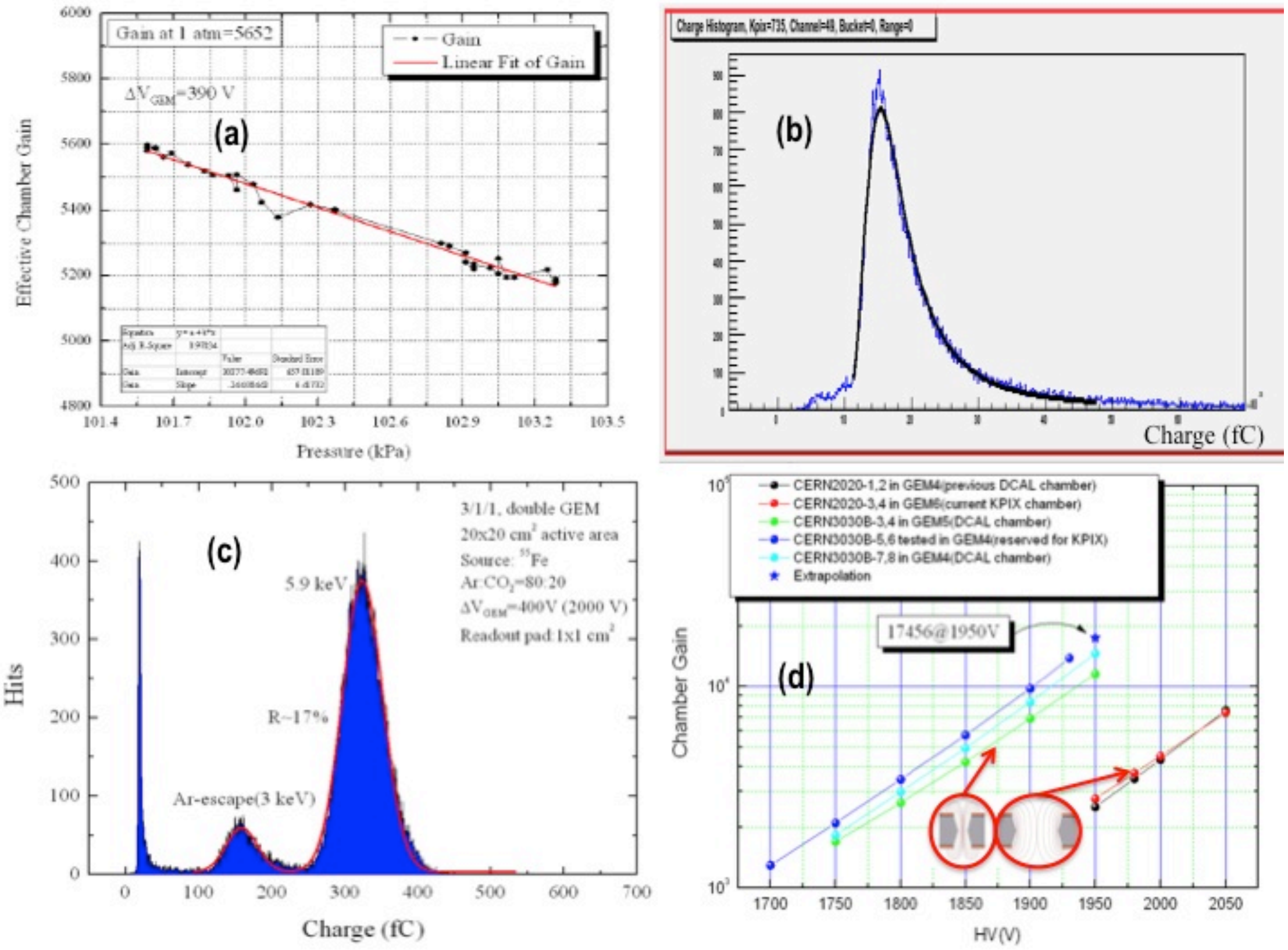

As a subsequent measurement to verify the functionality of the prototype detectors, we have characterized the prototype chambers on the bench. Figure 4.(a) shows $^{55}$Fe X-ray spectrum as a function of high voltage applied across the entire detector system. The spectra show expected behavior, the peak from the 5.9 KeV X-ray shifts to higher output charge values as the voltage increases, indicating the increased detector gain. Fitting aGaussian tothe 5.9KeV peak and comparing the resulting output charge to the expected number of ionization electrons in the drift gap, we obtained the gain of the chamber as a function of the potential difference across each of the two GEM foils in the detector as depicted in Fig.4.(b). The measured gain is consistent with other measurement of double GEM detectors, taking into account the gas composition.

We took cosmic ray data using two 2cmx3cm counters arranged perpendicular to each other but sandwiching the detector to ensure the center 2cmx2cm trigger area. Figure 4.(c) shows the scatter plot of the cosmic ray hit distributions, and Fig. 4.(d) shows the lego plot of the cosmic ray hit, clearly

**Figure 4. (a) Characteristic $^{55}$Fe X-ray peaks as a function of HV across the entire detector system (b) Measured detector gain obtained from $^{55}$Fe source runs as a function of potential across each GEM foil (c) Cosmic ray hit scatter plot of the 64 channel GEM detector. The trigger consisted of coincidences of two 2cmx3cm counters arranged perpendicular to each other, shadowing 2cmx2cm area in the center. (d) The lego plot of the cosmic ray hit map.**

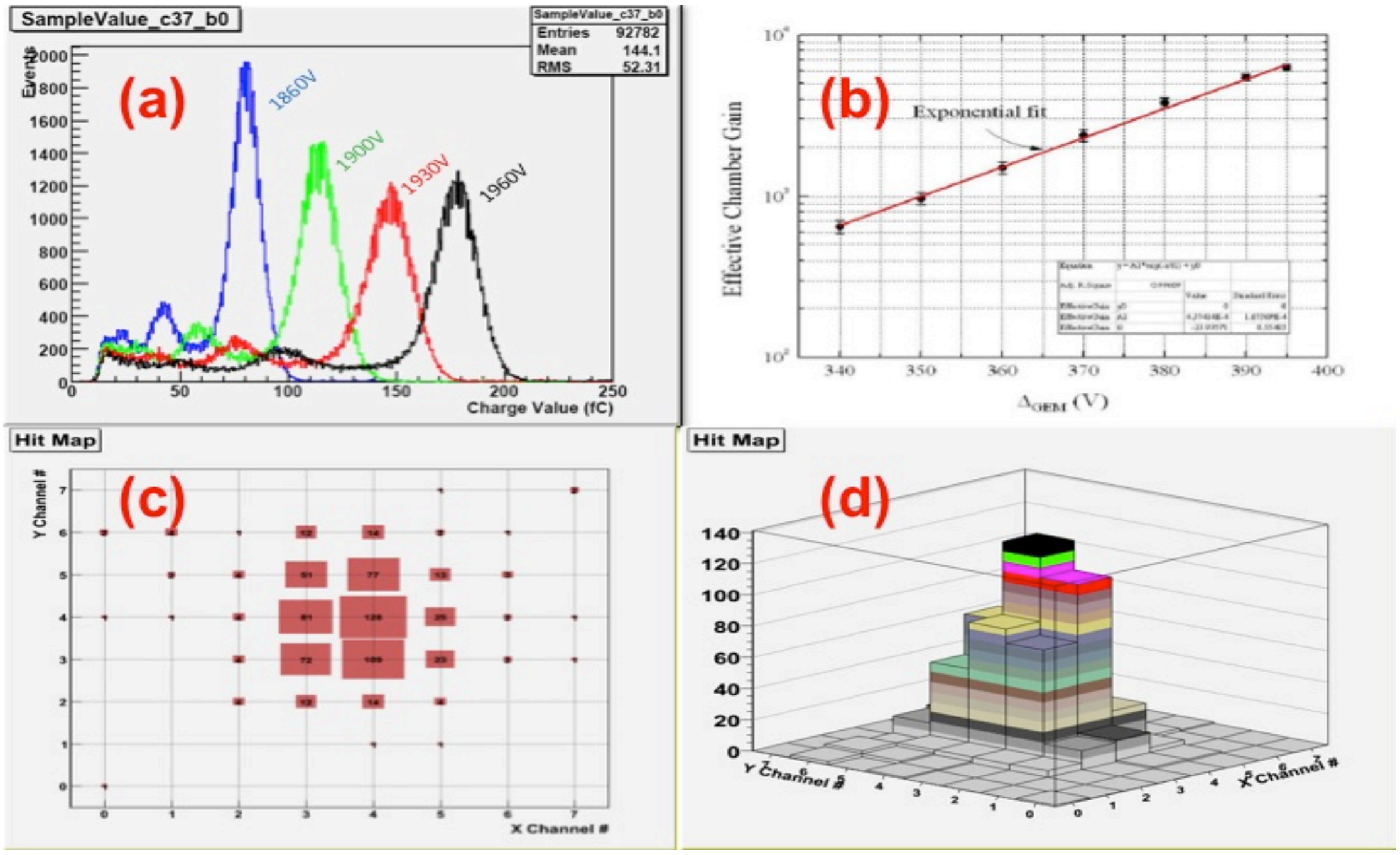

demonstrating the area shadowed by the trigger counters.

Unfortunately, in the process of preparing for beam tests originally planned for January 2011, the last KPiX 7 chip stopped functioning late 2010. We suspect the chip's demise was caused by sparks from the detector. This incident forced us to switch over to the 512 channel KPiX9 chip. This chip was integrated to the existing 64 active channel anode board to save cost. At the time of this presentation, we have characterized the new chip and were taking cosmic ray data with the beneficial collaboration between the UTA and SLAC teams.

**4. Integration with One Bit DCAL Chip**

Since ultimately the GEM detector will be used as the sensitive gap detector for a calorimeter in one bit readout format, it is of critical importance for the detector to be readout using the 1-bit readout electronics based on the DCAL chip developed by the Argonne National Laboratory and Fermi National Accelerator Laboratory teams. The DCAL chip has 64 readout channels and have been used to read out signals from a Resistive Plate Chamber (RPC) detector based DHCAL.

Figure 5.(a) shows a VME crate with various DCAL DAQ system components integrated into it. This system has the capability to expand to a large number of channels. Figure 5.(b) shows a 20cmx20cm anode board with 4 DCAL chips to read out a total of 256 channels interfaced to a 30cmx30cm dummy board with integration notches for gas tightness. This interfaced anode board is integrated into the 30cmx30 GEM prototype detector as shown in the photo.

We performed a noise scan of the boards to determine the optimal threshold for each of the two DCAL boards we have integrated with the chambers. The threshold turns are out to be on the order of 30 DAC counts which correspond to about 9 fC charge threshold. Once the threshold was determined we took radioactive source data as well as cosmic ray data with the DCAL readout system. Figure 5.(c) shows a lego plot of the hits from cosmic rays triggered by a 10cmx10cm scintillation counter trigger. Figure 5.(d) shows a contour plot of an X-ray image of a wrench taken with an elevated $^{55}Fe$ source that illuminated the entire detector with the wrench blocking part of the detector. Although it is not quite clear due to the coarse readout pad size of 1cmx1cm, one can make out the shape of the wrench. These exercises demonstrated the functionality of the DCAL readout system integrated with GEM prototype detectors. Given this, we built a total of three DCAL chambers to expose to particle beams in August 2011 at Fermilab Beam Test Facility.

**Figure 5. (a) DCAL DAQ system based on VME-PCI readout system. (b). 20cmx20cm DCAL readout board with 256 channel readout capability along with the GEM detector with full DCAL integration connected to the DCAL DAQ system (c) Lego plots of cosmic ray hits triggered with 10cmx10cm cosmic trigger (d) X-ray image of a wrench as a bench test after the integration of the DCAL electronics.**

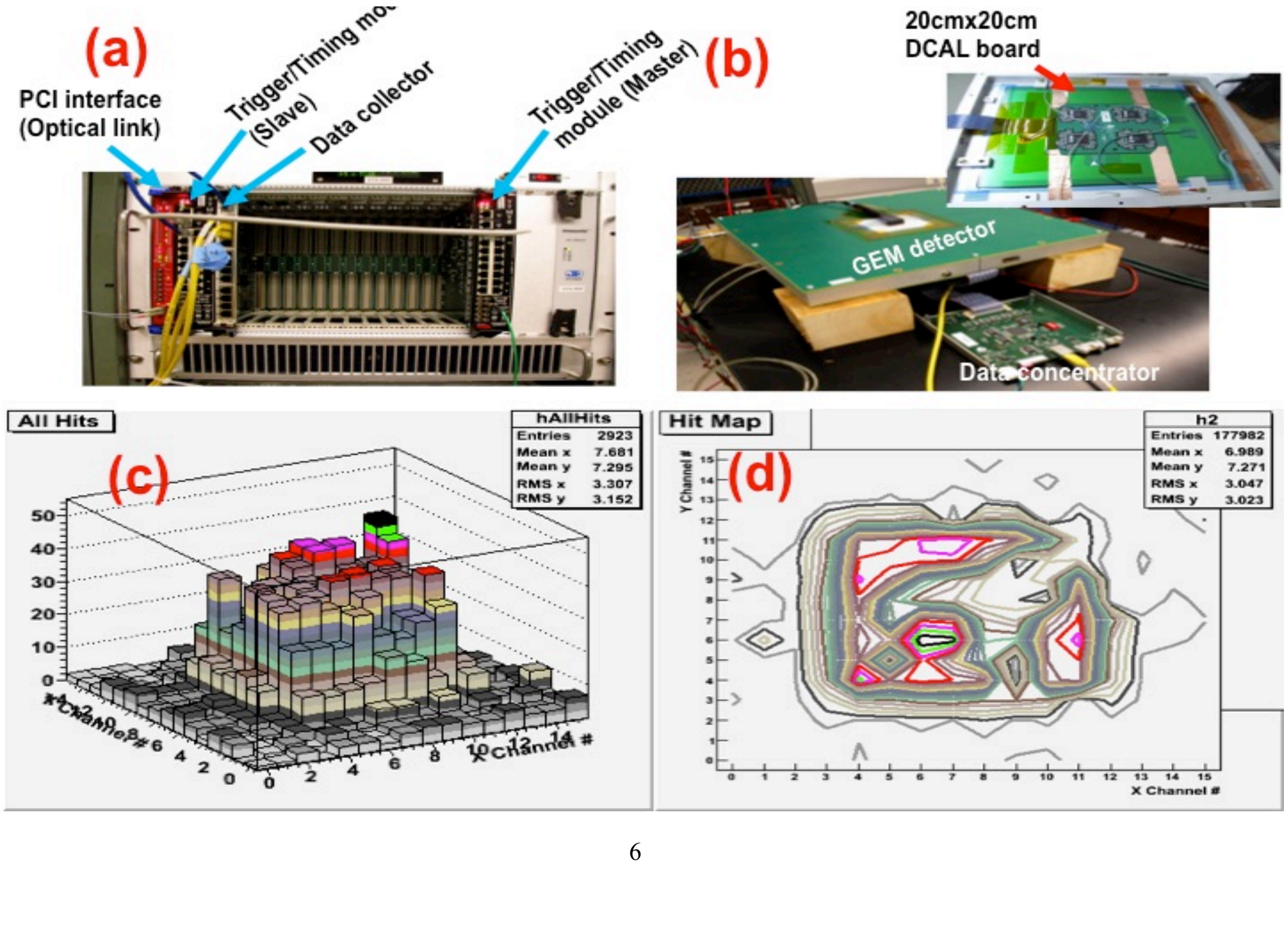

**5. Development of Large Scale GEM Detectors**

The next phase to characterization of GEM chambers is the development, characterization and construction of the full scale $1m^3$ GEM chambers for DHCAL. We plan on constructing a $1m\times 1m$ large scale chamber using three $1m\times 33cm$ unit chambers as shown in Fig.6.(a). Together with the CERN GDD workshop, we developed $1m\times 33cm$ GEM foils for unit chambers. While we had been working successfully with the 3M Inc., they have decided to close their micro-flex circuit division in late 2007. For this reason and for the fact that the CERN GDD workshop has been working on developing cost effective technology to produce large size GEM foils, we have decided to place work with CERN to produce $1m\times 33cm$ GEM foils.

Five of the first development samples were produced using a one-side etching technique and were delivered. Figure 6.(b) shows a photo of one of the $1m\times 33cm$ foils. These foils consist of 31 independent HV strips that run along the length of the foil. We developed a foil qualification procedure based on the leakage current on each of the 31 strips, the amount of time needed for each strip to saturate and the number of strips with saturation time less than 2000 seconds. We categorized the five foils in three different levels using the criteria described above: pass-high for foils with the qualification ready to build a detector, pass-med for foils with a small number strips that do not pass the criteria and fail for the foils that have large number of failed strips. Two of the foils passed with pass-high and the other two passed with pass-med while the last foil was not tested since it failed CERN GDD qualification.

**Figure 6. (a) A schematic diagram of a $1m\times 33cm$ unit chamber with base steel plate and two half chamber length anode boards. (b) A photo of a $1m\times 33cm$ GEM foil developed jointly with CERN GDD workshop. (c) Cross section of a $1m\times 33cm$ unit chamber (d) A schematic diagram of a $1m\times 1m$ plane that consists of three unit chambers**

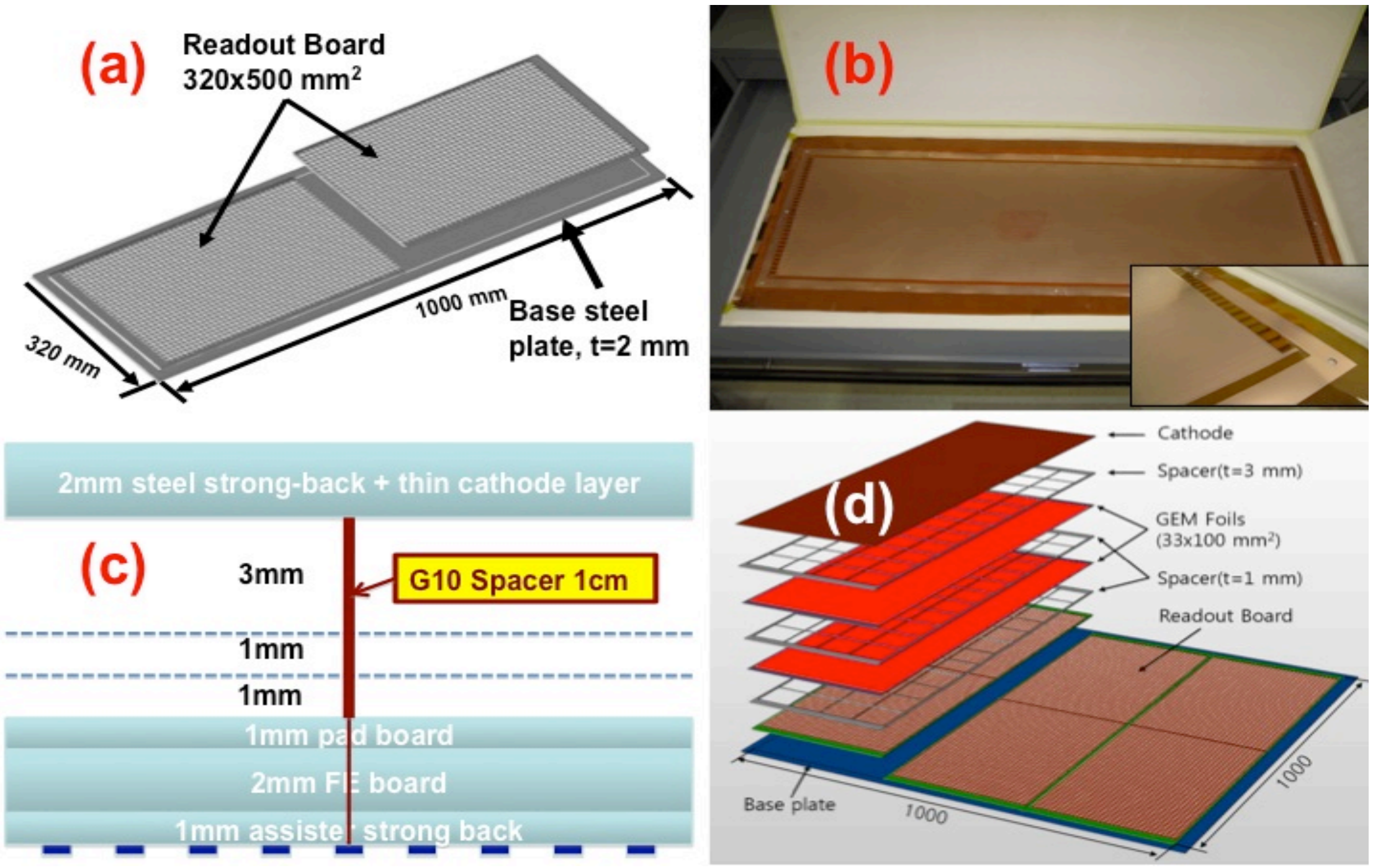

Figure 6.(c) shows a cross section of a $1\,\mathrm{m}\times 33\,\mathrm{cm}$ unit chamber. We originally planned on using a large scale G10 spacer with a 1cm bar in the center to provide a mechanical support to the two 32cmx50cm anode boards. This will maintain the mechanical integrity of the chamber and gas tightness. Through the discussion with the ANL team, we started considering direct gluing of the two boards on the edges. We will have to test and make sure that this scheme provides sufficient mechanical integrity and gas tightness. Three of these unit $1\,\mathrm{m}\times 33\,\mathrm{cm}$ chambers will make up a $1\,\mathrm{m}\times 1\mathrm{m}$ detector plane, using a $1\,\mathrm{m}\times 1\mathrm{m}$ steel base plate as shown in Fig.6.(d).

## 6. Future Plans

We will develop the mechanical structure, the electronic readout board schemes and the schemes for connecting the three unit chambers into one $1\mathrm{m}\times 1\mathrm{m}$. We plan to construct these unit chambers through 2013. Through the construction stage, we will test these unit chambers with source, cosmic rays and with beams. We will then put minimum five of these $1\mathrm{m}\times 1\mathrm{m}$ chambers when possible into the CALICE [13] beam test calorimeter stack in mid 2014 to late 2015. The primary goals of this beam test are to measure the responses and energy resolution of a GEM-based DHCAL. This result should be compared to that of a DHCAL using RPCs and other analog HCALs. This full scale prototype will be tested jointly with CALICE Si/W or Scintillator/W ECAL and a tail catcher (TCMT), using the CALICE mechanical support structure which used in many previous beam tests.

Finally, we have been considering an alternative to the standard GEM foil technology. The "Thick-GEM" (TGEM) [14], [15] has been shown that it can, in a single layer, achieve multiplication levels typical of at least a double-GEM device. A TGEM is a printed circuit board, clad with copper on both sides through which holes have been drilled. A typical configuration might be a 0.4 mm thick board with 0.3 mm diameter holes spaced 1 mm apart. We are collaborating with the Weizmann Institute to work on production and prototyping of GEM chambers using large scale TGEMs. TGEM team also has made significant progress and presented their results at this conference.

## 7. Conclusions

The UTA HEP group has made significant progress in the development and understanding of $30\,\mathrm{cm}\times 30\,\mathrm{cm}$ double GEM chambers. We have successfully integrated these prototype chambers with the 13bit KPiX analog readout board as well as one bit DCAL readout boards. We have characterized the latest 512 channel KPiX9 chips and have characterized the GEM chamber on the bench using various radioactive sources and cosmic rays. We plan on conducting beam tests using four $30\,\mathrm{cm}\times 30\,\mathrm{cm}$ chambers, one with 13-bit KPiX and three with 1-bit DCAL readout system. We have qualified the first set of 5 large GEM foils of dimension $33\,\mathrm{cm}\times 1\mathrm{m}$. We have completed the design of G10 spacers for the unit chamber and are preparing a clean room for foil certification and chamber construction. We are working on mechanical design for $33\,\mathrm{cm}\times 1\mathrm{m}$ unit chambers and $1\mathrm{m}\times 1\mathrm{m}$ DHCAL plane. Finally, we are working with Weizmann Institute for using large scale TGEM for DHCAL.

## Acknowledgements

The support for work is partially provided by the U.S. Department of Energy and the U.S. National Science Foundation. The authors also express appreciation to the KPiX development team at the Stanford Linear Accelerator Center and the University of Oregon for making modification on the KPiX readout chip and for working closely in integration of the chip with prototype GEM chambers. We also would like to express our gratitude for Argonne National Laboratory team for helping us integrating DCAL readout system with our chambers.